\documentclass[sigconf]{acmart}

\renewcommand\footnotetextcopyrightpermission[1]{} 
\usepackage{hyperref}
\usepackage{algorithm}
\usepackage{algpseudocode}

\AtBeginDocument{%
  }

\begin{document}

\title{Orthogonal Low Rank Embedding Stabilization}

\author{Kevin Zielnicki}
\email{kzielnicki@netflix.com}
\author{Ko-Jen Hsiao}
\email{khsiao@netflix.com }
\affiliation{%
  \institution{Netflix}
  \city{Los Gatos}
  \state{California}
  \country{USA}
}

\begin{abstract}
The instability of embedding spaces across model retraining cycles presents significant challenges to downstream applications using user or item embeddings derived from recommendation systems as input features. This paper introduces a novel orthogonal low-rank transformation methodology designed to stabilize the user/item embedding space, ensuring consistent embedding dimensions across retraining sessions. Our approach leverages a combination of efficient low-rank singular value decomposition and orthogonal Procrustes transformation to map embeddings into a standardized space. This transformation is computationally efficient, lossless, and lightweight, preserving the dot product and inference quality while reducing operational burdens. Unlike existing methods that modify training objectives or embedding structures, our approach maintains the integrity of the primary model application and can be seamlessly integrated with other stabilization techniques.
\end{abstract}

\maketitle

\section{Introduction}

Many recommender models rely on internal embeddings to model users and items as an intermediate step prior to the final inference output \cite{Koren2009MatrixFT, Rendle2010Factorization}. These embeddings encapsulate the model's understanding of the users and items in a fixed-dimensional vector representation \cite{Bengio2013Representation}. The same properties that make embedding representations useful for the recommender model can make them useful for a variety of downstream purposes, such as input features to other models or for semantic similarity \cite{Steck2024Similarity}. Embeddings are also foundational to Retrieval-Augmented Generation (RAG), enabling Large Language Models (LLMs) to retrieve semantically relevant information from large corpora and integrate it into their generation process \cite{Lewis2020, Borgeaud2022}. However, the embedding space has arbitrary and uninterpretable dimensions, and most problematically for embedding consumers, it is completely incompatible from one model training run to the next \cite{Hu2022BackwardComp}.

In an industrial recommendation setting, the instability of embedding spaces poses significant challenges \cite{Zheng2025Stable}. Downstream consumers must tightly integrate with the retraining and redeployment cycles of the base recommender model, risking the introduction of complex bugs when assumptions about the embedding space's structure are invalidated with each retraining \cite{Hu2022BackwardComp}. Typically, downstream models that use user and item embeddings from the base model can retrain alongside it to avoid incompatibility issues. However, consider a scenario where a downstream application samples 10 million users daily, generates user features, and uses the base model to produce user embeddings. The application model undergoes daily retraining, utilizing the past three months' worth of daily generated features and embeddings as its training data, which amounts to approximately 900 million user embeddings. If the base model is retrained from scratch daily, regenerating or backfilling all these 900 million embeddings daily becomes infeasible, slow or costly, especially if the base model's complexity or the volume of user interaction data is high \cite{Rangadurai2022NxtPost}.

To address this, one might consider enforcing embedding similarity across retraining runs through additional regularization \cite{Shen2020BackwardComp}. However, this approach can reduce training efficiency and still does not ensure full compatibility of embedding dimensions across different retraining sessions. Alternatively, reducing the retraining frequency of the base model to, say, once a month could minimize the need for frequent backfilling. Yet, this approach risks degrading embedding quality over time as user behavior evolves \cite{mahadevan2023costeffectiveretrainingmachinelearning, Verachtert2023Scheduling}. While daily fine-tuning or continuous training of the base model could maintain high quality, it reintroduces the challenge of dealing with different or updated base models daily. Therefore, stabilizing the embedding space becomes crucial, ensuring consistent embedding dimensions and reducing the operational burden on downstream applications.

This work describes an orthogonal low-rank transformation methodology for stabilizing the user/item embedding space, so that downstream consumers can rely on the embedding dimensions having a consistent meaning even as the base recommender model is retrained and redeployed. This transformation works for any factorization-style model which defines scores as a dot product between user and item embeddings. In addition, it extends naturally to softmax-based neural network models, where the final layer is a softmax over item logits. In this setting, the penultimate layer output can be interpreted as the user embedding, while each column of the softmax (or final feedforward) weight matrix corresponds to an item embedding—assuming no bias term is present. Additionally, the orthogonal low-rank transformation has several desirable properties:
\begin{enumerate}
    \item Fast. The transformation uses a simple matrix product that is efficient to compute, so overhead is negligible.
    \item Lossless. The transformation preserves the dot product, so inference is unchanged and no information is lost.
    \item Lightweight. The transformation is learned by a small number of calls to well-optimized linear algebra functions, not a model training loop.
    \item Intuitive. The transformation uses a singular value decomposition which transforms the embedding space into a principal components representation, which is more interpretable than a raw embedding space.
\end{enumerate}

Compared to methods that learn backward compatible embeddings and transformations \cite{Hu2022BackwardComp} or different embedding structure \cite{Zheng2025Stable}, our approach does not distort the primary model application and does not need additional training processes. Unlike methods that require additional backward compatibility learning procedures \cite{Shen2020BackwardComp}, our approach inherently utilizes the structure of the recommendation problem, eliminating the need for extra steps in future model training. Finally, as a lightweight post-processing step, our method can be easily combined with other stabilization approaches. Most other methods aim to minimize changes in embeddings or maintain consistent inner product results or rankings. However, these do not "freeze" the meaning of each dimension. Our approach selects one embedding as the reference and transforms all others into this reference's transformed space.

\section{Methodology}

\subsection{Summary}

The orthogonal low-rank transformation has two main components: an efficient low rank singular value decomposition of the score space (Section \ref{low-rank-svd}), followed by an orthogonal Procrustes transformation (Section \ref{orthogonal-procrustes}) to a fixed SVD-transformed reference space. Taken together, these two components yield a single matrix transformation that maps embeddings from any model training run into a consistent standardized space.

\subsection{Low Rank SVD} \label{low-rank-svd}

Factorization-style models work by learning low-rank approximations of a high-dimensional score space. Suppose we have a score matrix $X$ consisting of $n$ items and $m$ users. If we select an embedding dimension $e$, we can represent items as an $n \times e$ matrix $T$ and users as an $m \times e$ matrix $W$, such that $X \approx TW^T$.

Generally, the embedding dimension is arbitrarily ordered and no specific meaning is attributed to the 1st vs. the $i$th element. This degeneracy contributes to embedding space instability, as small perturbations in initial conditions can lead to very different low-rank approximations. In contrast, singular value decomposition provides a low-rank approximation which is unique up to sign flips, getting us closer to a stable space (plus other benefits). However, SVD operates in $O(mn^2)$ time, which is not tractable to compute for a large score matrix. To address this, the recommender systems community has developed several scalable variants of SVD—such as incremental SVD and randomized SVD \cite{sarwar2002incremental,halko2011finding}—which have proven effective for factorization-based models. However, these methods are not well-suited for softmax-style neural network architectures, where embeddings emerge implicitly from weight matrices and activations, and where the score matrix itself is not explicitly formed or factorized.

Fortunately, we can leverage the existing low-rank approximation $ TW^T $ to efficiently compute the SVD on a smaller $ e \times e $ matrix $ R_T R_W^T $. This is achieved by using the QR decomposition of $ T $ and $ W $, where $ R_T $ and $ R_W $ are the $ R $ matrices from the QR factorization of $ T $ and $ W $, respectively. This method scales only linearly in $m$ and $n$, and is practical to compute at machine learning scale.

In practice, we need not just the transformed space, but also the transformation into the new space. We can achieve this by computing transformation matrices $M_T$ and $M_W$ such that $T M_T = U S^{1/2}$ and $W M_W = V S^{1/2}$, using matrix identities to avoid taking the inverse of R when computing $M_T$ and $M_W$. This yields transformations that preserve the matrix product, $T M_T (W M_W)^T = USV = TW^T$ (Algorithm \ref{low-rank-svd-algo2}).

\begin{algorithm}
\caption{Low-Rank SVD Transformation of $TW^{T}$} \label{low-rank-svd-algo2}
\begin{algorithmic} 
\Procedure{LowRankSVDTrans}{$T$,$W$}
    \State $Q_T,R_T \gets \mathtt{qr}(T)$
    \State $Q_W,R_W \gets \mathtt{qr}(W)$
    \State $U_R,S,V_R^T \gets \mathtt{svd}(R_T R_W^T)$
    \State $M_T \gets R_W^T V_R S^{-1/2}$
    \State $M_W \gets R_T^T U_R S^{-1/2}$
    \State \textbf{return} $M_T, M_W$ 
\EndProcedure
\end{algorithmic}
\end{algorithm}

\subsection{Orthogonal Procrustes} \label{orthogonal-procrustes}

The low-rank SVD from the previous section only partly fulfills our goal. It can provide a standardized non-degenerate factorization of the embedding space for a given model training run, but separate training runs can have sufficiently different variance structure such that the SVD of each day's score matrix will still not produce a compatible space.

We will address this by choosing a fixed day to define a standardized space, and then learning a mapping from other days into that standardized space. This mapping could be an arbitrary matrix, but for numerical stability and computation efficiency, it is convenient to restrict the mapping to be an orthogonal matrix. In other words, given matrices $A$, $B$, we wish to find an orthogonal matrix $R$ such that $RA \approx B$. This is known as the Orthogonal Procrustes problem, and it has an efficient known solution which is Lipschitz continuous and thus stable for small input perturbations.

\begin{algorithm}
\caption{Orthogonal Procrustes} \label{orthogonal-procrustes-algo}
\begin{algorithmic} 
\Procedure{OrthoProcrustes}{$A$,$B$}
    \State $M \gets B^T A$
    \State $U, S, V^T \gets \mathtt{svd}(M)$
    \State $R \gets U V^T$
    \State \textbf{return} $R^T$ 
\EndProcedure
\end{algorithmic}
\end{algorithm}

\subsection{Stabilizing Transformation} \label{stab-transform}

With the building blocks of the previous two sections, we can now construct the full orthogonal low rank stabilizing transformation. First, we'll select a reference model training run indexed as $k=0$ to seed our standard space, and further runs indexed as $k=1...i$ to stabilize. Let $T_k$ and $W_k$ be the user and item embeddings from run index $k$, and $M_{Tk}$ and $M_{Wk}$ be the results of $\mathtt{LowRankSVDTrans}(T_k,W_k)$. Further, let $T'_k = T_k M_{Tk}$ be the SVD transformed item embeddings. Then, let $R_k = \mathtt{OrthoProcrustes}(T'_k, T'_0)$. This gives the final transformations $M'_{Tk} = M_{Tk}R_k$ and $M'_{Wk} = M_{Wk}R_k$ such that the stabilized item and user embeddings are $\hat{T}_k = T_k M'_{Tk}$ and $\hat{W}_k = W_k M'_{Wk}$ respectively.\footnote{Notice that we define the map $R_k$ using only the items. It would be possible to use the users, or both the users and items, or a random subset of either. The choice of items-only here is because we generally expect items to be more stable than users.}

Note that because of orthogonality, the score matrix product $\hat{T}\hat{W}^T = TW^T$ is preserved! Also, the transformations are small $e\times e$ matrices which are easy to store and retrieve, and add only a single matrix product to embedding inference, which is computationally negligible compared to the rest of the neural network. (This operation can be added to the neural network model as an additional feedforward layer.)

\section{Application}

In Section \ref{stab-transform}, we discuss selecting a reference model training run, indexed as $k=0$, to establish our standard space and learn the transformation for each subsequent run $k=1...i$. This requires specially storing the item embeddings from the $k=0$ run for future reference. However, treating any run as a special case and passing this information to all future runs can be prone to errors. Fortunately, there is a solution. For the $k=1$ run, we transform the embeddings to the standardized space defined by the  $k=0$ run. Once this transformation is complete, the  $k=1$ run's stabilized embeddings are also in the standardized embedding space. Therefore, instead of referencing the  $k=0$ run, the  $k=2$ run can simply reference the stabilized embeddings from the previous $k=1$ run ($\hat{W}_k$) to learn the transformation. With a fixed vocabulary, this is exactly equivalent to using the $k=0$ run, due to orthonormality.

In an industrial setting, user embeddings can be generated either through offline batch processing, an event-triggered approach or live/online inference. In offline batch processing, a pipeline loads the latest base model and runs inference for all users to generate their embeddings. In this scenario, the transformation, which involves multiplying a small $e\times e$ matrix with all user embeddings, can be efficiently executed using Spark or other big data processing tools. In the event-triggered approach, a user's embedding is updated whenever significant interaction events occur, or alternatively, after a set number of events. In this case, the $e\times e$ matrix multiplication can be integrated into the base model itself as an additional neural network layer and the model would produce the stabilized embeddings directly. In fact, this trick can also be applied to the offline batch processing use case.

\section{Validation}

To validate our method, we compare user and item embeddings from a transformer-based recommender model trained on proprietary data, using three datasets from different time periods: (A) $T=0$, (B) $T=+2$ weeks, and (C) $T=+4$ weeks. We use three different comparisons shown in table \ref{results-table}: average same-user cosine similarity, average same-item cosine similarity, and rank correlation (using rank-biased overlap).

\begin{table}[ht]
    \centering
    \caption{Comparison of User and Item Embeddings Across Different Time Periods Relative to Period A}
    \label{results-table}
    \begin{tabular}{lcccccc}
        \toprule
        \textbf{Metric} & \textbf{Period} & \textbf{Unstabilized} & \textbf{Stabilized} \\
        \midrule
        User Similarity & A vs B & 0.013 & 0.82 \\
        User Similarity & A vs C & 0.014 & 0.75 \\
        Item Similarity & A vs B & 0.042 & 0.82 \\
        Item Similarity & A vs C & 0.095 & 0.82 \\
        Rank Correlation & AA vs AB & 0.035 & 0.62 \\
        Rank Correlation & AA vs AC & 0.041 & 0.53 \\
        \bottomrule
    \end{tabular}
    \label{tab:comparison}
\end{table}

For cosine similarity comparisons, we computed the average cosine similarity between users/items in period A and the same users/items in periods B and C. Without stabilized embeddings, different periods do not share a consistent basis, so similarities are close to 0. With stabilization, similarities are 0.75 or higher, demonstrating a meaningful consistency of user and item representations across different periods.

For rank correlation, we compare user-item scores computed from user and item representations both from period A, against user-item scores computed with user representations from period B/C and item representations from period A. Without stabilization, user-item scores using representations from different days are uncorrelated with user-item scores using same-day representations, while stabilization yields correlations > 0.5.

\section{Author Bios}

Kevin Zielnicki is a research scientist/engineer at Netflix. His work focuses on recommendation systems and leveraging data products derived from recommendation systems to address data science challenges. He earned his PhD from the University of Illinois at Urbana-Champaign, focusing on quantum information processing.

Ko-Jen Hsiao is a research scientist/engineer at Netflix working on recommendation systems. He has expertise in applying machine learning at scale, constructing and A/B testing ranking algorithms and recommendation systems. Currently, he focuses on foundational models for Netflix's personalized experiences. He earned his PhD from the University of Michigan, where he specialized in combining disparate information for machine learning applications. 

\begin{acks}

We sincerely thank colleagues Nathan Kallus and Justin Basilico for their invaluable feedback during the internal review of our paper draft. Their insights and suggestions significantly enhanced the quality of our work.

\end{acks}

\bibliographystyle{ACM-Reference-Format}
\bibliography{reference}

\end{document}